\documentclass[onecolumn]{revtex4}

\usepackage{graphicx}% Include figure files
\usepackage{dcolumn}% Align table columns on decimal point
\usepackage{bm}% bold math
\usepackage{color}
\usepackage{amssymb,amsmath}

\input epsf

\begin{document}

\title{\Large{Natures of Statefinder Parameters and {\it Om} Diagnostic for Cardassian Universe in Ho$\check{\text r}$ava-Lifshitz Gravity}}

\author{\bf Piyali Bagchi Khatua$^1$\footnote{piyali.bagchi@yahoo.co.in}
and Ujjal Debnath$^2$\footnote{ujjaldebnath@yahoo.com} }

\affiliation{$^1${Department of Computer Sc and Engg, Netaji Subhas Engineering College, Garia, Kolkata-700 152, India.\\
$^2$Department of Mathematics, Bengal Engineering and Science
University, Shibpur, Howrah-711 103, India.} }

\date{\today}

\begin{abstract}
In this work, we have considered Cardassian Universe in
Ho$\check{\text r}$ava-Lifshitz gravity. Four types of Cardassian
Universe models i.e., polytropic/power law, modified polytropic,
exponential and modified exponential models have been considered
for accelerating models. The natures of statefinder parameters,
deceleration parameter, $Om$ diagnostic and EoS parameters have
been investigated for all types of Cardassian models in
Ho$\check{\text r}$ava-Lifshitz gravity.
\end{abstract}


\maketitle
\section{\normalsize\bf{Introduction}}

Recent observations of the luminosity of type Ia supernovae
indicate an accelerated expansion of the universe [1-7] and lead
to the search for a new type of matter, called dark energy, which
violates the strong energy condition [8-10]. For explaining the
accelerating expansion of the universe, large number of
cosmological models have been proposed. Dark energy model is
proposed by assuming an energy component with negative pressure in
the universe, this dark energy dominates the total energy density
of the universe and drives its acceleration of expansion at recent
times. Some authors have explored possible explanations for the
acceleration: (i) cosmological constant, (ii) quintessence
[11-17], (iii) gravitational leakage into extra dimensions
[18-19]. Accelerated expansion also depends on other observations
such as the cosmic microwave background (CMB) [20] and galaxy
power spectra [21]. Another reason of acceleration is the
geometric effect of the general relativity fails in the present
cosmic time space scale. The model proposed by Freese and Lewisan,
called Cardassian model [22-26], assume that the universe is flat
and accelerating which consists only of matter and radiation [27].
The geometry is flat as required by measurements of the cosmic
background radiation, so that there are no curvature terms in the
equation and no vacuum term in the equation and so the model does
not address the cosmological constant. In Cardassian Model, we
take $g(\rho)$ as a function of $\rho$ that returns
simply to $l^2\rho/3$ at early epochs, where $l^2=8\pi G_c$.\\

 On the basis of the recent observation one can
state that if Einstein's theory of gravity is acceptable on
cosmological scales, then our universe must dominated by a
mysterious form of energy called dark energy. Recently
Ho$\check{\text r}$ava [28-29] proposed a new theory of gravity,
which renormalizable with higher spatial derivatives in four
dimensions which reduces to Einstein's gravity with non vanishing
cosmological constant in IR but with improved UV behaviours. In
Lifshitz [30] scalar field theory the time dimension has weight 3
if a space dimension has weight 1 and this theory is called
Ho$\check{\text r}$ava-Lifshitz gravitational theory.
Ho$\check{\text r}$ava-Lifshitz gravity has been studied and
extended in detail and applied as a
cosmological framework of the universe [31-33].\\

As so many cosmological models have been developed, so for
discrimination between these contenders, Sahni et al [34] proposed
a new geometrical diagnostic named the statefinder pair $\{r,s\}$,
where $r$ is generated from the scalar factor $a$ and its
derivatives with respect to the cosmic time $t$ and $s$ is a
simple combination of $r$ and the deceleration parameter $q$.
Clear differences for the evolutionary trajectories in the $r$-$s$
plane have been found. In this work, we have discussed four
different types of Cardassian Universe models in Ho$\check{\text
r}$ava-Lifshitz gravity. In every case we find the statefinder
parameters, deceleration parameter and $Om$ diagnostic [34-36].\\

\section{\bf{Cardassian Universe in Ho$\check{\text r}$ava-Lifshitz Gravity}}

 The Arnowitt-Deser-Misner formalism of the full metric is written as [37],

\begin{equation}
ds^{2}=-N^{2}dt^{2} + g_{ij}(dx^{i} + N^{i}dt)(dx^{j} + N^{j}dt)
\end{equation}

Under the detailed balance condition the full action condition of
Ho$\check{\text r}$ava-Lifshitz gravity is given by,

\begin{eqnarray*}
S=\int dt d^{3}x \sqrt{g}N \left[\frac{2}{\kappa^{2}}(K_{ij}K^{ij}
- \lambda K^{2}) + \frac{\kappa^{2}}{2 \omega^{4}}C_{ij}C^{ij}-
\frac{\kappa^{2}\mu \epsilon^{ijk}}{2 \omega^{2} \sqrt{g}}
R_{il}\nabla_{j}R^{l}_{k}\right.
\end{eqnarray*}
\begin{equation}
\left.+ \frac{\kappa^{2}\mu^{2}}{8}R_{ij}R^{ij}+
\frac{\kappa^{2}\mu^{2}}{8(3 \lambda - 1)}\left(\frac{1 -4
\lambda}{4}R^{2} + \Lambda R - 3 \Lambda^{2} \right)\right]
\end{equation}

where the extrinsic curvature and Cotton tensor is defined as,
$K_{ij}=\frac{1}{2N}(\dot{g}_{ij} - \nabla_{i}N_{j}-
\nabla_{j}N_{i})~~~and~~~~~~C^{ij}=\frac{\epsilon^{ikl}}{\sqrt{g}}\nabla_{k}(R^{j}_{i}
- \frac{1}{4}R \delta^{j}_{l})$. The covariant derivatives are
defined w.r.t. the spatial metric $g_{ij}$. $\epsilon^{ijk}$ is
the totally antisymmetric unit tensor, $\lambda$ is a
dimensionless coupling constant and the variable $\kappa$ ,
$\omega$ and $\mu$ are constants with mass dimensions $-1,~ 0,~ 1$
respectively. Also $\Lambda$ is a positive constant, which as
usual is related to the cosmological constant
in the IR limit.\\

Now, in order to focus on cosmological frameworks, we impose the
so called projectability condition and use a
Friedmann-Robertson-Walker (FRW) metric we get, $N=1,
g_{ij}=a^{2}(t)\gamma_{ij},      N^{i}=0 $ with $\gamma_{ij}
dx^{i} dx^{j}= \frac{dr^{2}}{1 - kr^{2}}+ r^{2}d\Omega^{2}_{2}$
where $k=0, -1, +1$ corresponding to flat, open and closed
respectively. By varying $N$  and $g_{ij}$, we obtain the
non-vanishing equations of motions:

\begin{equation}
H^{2}=\frac{\kappa^{2}}{6(3\lambda -1)} ~ \rho +
\frac{\kappa^{2}}{6(3\lambda -1)} \left[\frac{3 \kappa^{2} \mu^{2}
k^{2}}{8(3\lambda -1)a^{4}} + \frac{3 \kappa^{2} \mu^{2}
\Lambda^{2}}{8(3\lambda -1)}\right] - \frac{ \kappa^{4} \mu^{2}k
\Lambda }{8(3\lambda -1)^{2}a^{2}}
\end{equation}
and
\begin{equation}
\dot{H} + \frac{3}{2}H^{2}= - \frac{\kappa^{2}}{4(3\lambda -1)}~p
- \frac{\kappa^{2}}{4(3\lambda -1)} \left[\frac{ \kappa^{2}
\mu^{2} k^{2}}{8(3\lambda -1)a^{4}}- \frac{3 \kappa^{2} \mu^{2}
\Lambda^{2}}{8(3\lambda -1)}\right] - \frac{ \kappa^{4} \mu^{2}k
\Lambda }{16(3\lambda -1)^{2}a^{2}}
\end{equation}

where $H\equiv\frac{\dot{a}}{a}$ is the Hubble parameter. Here
$G_c$ and $G$ is defined as, $G_{c}=\frac{\kappa^{2}}{16
\pi(3\lambda -1)}$ and $G=\frac{\kappa^{2}}{32 \pi}$ where $G_{c}$
is the ``cosmological" Newton's constant and $G$ is the
``gravitational" Newton's constant.\\

We can re-write the above equations as [38],
\begin{equation}
H^{2} + \frac{k}{a^{2}}=\frac{l^2 \rho}{3} + \frac{k^{2}}{2
\Lambda a^{4}} + \frac{\Lambda}{2}
\end{equation}
and
\begin{equation}
\dot{H} + \frac{3}{2}H^{2} + \frac{k}{2a^{2}}= -\frac{l^2 p}{2} -
\frac{k^{2}}{4 \Lambda a^{4}} + \frac{3\Lambda}{4}
\end{equation}

Here, $\rho$ and $p$ are respectively the energy density and
pressure of the universe, $l^2=8\pi G_c$ and choosing $8\pi G=1$.\\

Freese and Lewis [22] constructed Cardassian universe models so
that, in Cardassian models the universe is flat and accelerating,
and yet contains only matter (baryonic or not) and radiation. The
above equation governing the expansion of the universe is modified
to,
\begin{equation}
H^{2} + \frac{k}{a^{2}}=\frac{l^2 g(\rho)}{3} + \frac{k^{2}}{2
\Lambda a^{4}} + \frac{\Lambda}{2}
\end{equation}
which gives,
\begin{equation}
H=\sqrt{-\frac{k}{a^{2}}+\frac{l^2 g(\rho)}{3} + \frac{k^{2}}{2
\Lambda a^{4}} + \frac{\Lambda}{2}}
\end{equation}

where $\rho$ is the total energy density of matter and radiation
and we will neglect the contribution of radiation for the
late-time evolution of the universe. The function $g(\rho)$
reduces to $\rho$ in the early universe. Now,
\begin{equation}
g(\rho)=\rho_{m}+\rho_{c}~~~~ and ~~~~~~p=p_m+p_c
\end{equation}
where $\rho_m$ and $\rho_c$ are the energy densities of matter and
Cardassian term of the universe respectively and $p_m$ and $p_c$
are the pressure of matter and Cardassian term of the universe
respectively. For any suitable Cardassian model, the following
requirements that should be fulfilled. (i) The function $g(\rho)$
should returns to the usual form of $\rho$ at early epochs in
order to recover the thermal history of the standard cosmological.
(ii) $g(\rho)$ should takes a different form at late times when
$z\thicksim \textrm{O}(1)$ in order to drive an accelerated
expansion as indicated by the observation of SNeIa. (iii) The
classical solution of the expansion should be stable, and the
sound speed $c_s^2$ of classical perturbations of the total
cosmological fluid around homogeneous FRW solutions cannot be
negative.\\

In order to guarantee the classical solution of the expansion is
stable, the sound speed $c_s^2$ of classical perturbations of the
total cosmological fluid around homogeneous FRW solutions should
always be greater than zero. If the expansion of the universe is
adiabatic, the sound speed of total cosmological fluid can be
represented by,
\begin{equation}
c_s^2=\frac{\partial p}{\partial \rho}
\end{equation}
Now the matter conservation equation gives,
\begin{equation}
\dot{\rho}_{m}+3H(\rho_{m}+p_m)=0
\end{equation}
and total fluid conservation equation gives,
\begin{equation}
\dot{\widehat{g(\rho)}}+3H(g(\rho)+p)=0
\end{equation}
From the above two equations we get,
\begin{equation}
\dot{\rho_{c}}+3H(\rho_{c}+p_c)=0
\end{equation}
From where we get,
\begin{equation}
p_c=(\rho_m+p_m)\frac{\partial
g(\rho)}{\partial\rho_m}-g(\rho)-p_m
\end{equation}
and
\begin{equation}
p=p_m+p_c=(\rho_m+p_m)\frac{\partial
g(\rho)}{\partial\rho_m}-g(\rho)=\rho_m(1+w_m)\frac{\partial
g(\rho)}{\partial\rho_m}-g(\rho)
\end{equation}
Now for dark matter $p_m=w_m \rho_m$, combining this with (11) we
get,
\begin{equation}
\rho_{m}=\rho_0 a^{-3(1+w_m)}~~~and~~~~p_{m}=\rho_0 w_m
a^{-3(1+w_m)}
\end{equation}
where $\rho_0$ be the integrating constant. Now the relation
between scale factor $a$ and the redshift $z$ is given by $a
=1/(1+z)$ i.e. $z=(1/a)-1$, and we replace all $a$ by $z$, where
value of $z$ is taken $z\ge -1$.

\section{\normalsize\bf{Statefinder Diagnostics, deceleration Parameter and} $Om$ \normalsize\bf{Diagnostic}}

The flat Friedmann model which is analyzed in terms of the
statefinder parameters. The trajectories in the $\{r, s\}$ plane
of different cosmological models shows different behavior. The
statefinder diagnostic of SNAP observations used to discriminate
between different dark energy models. The statefinder diagnostic
pair is constructed from the scale factor $a(t)$. The statefinder
diagnostic pair is denoted as $\{r,s\}$ and defined as [34],

\begin{equation}
 r=\frac{\dddot{a}}{a H^3}
 ~~~\text{and}~~~
s=\frac{r-1}{3(q-\frac{1}{2})}
\end{equation}

where $q$ is the deceleration parameter given by, $q=-\frac{a
{\ddot{a}}}{{\dot{a}}^2}$. If universe is present with dark matter
then the parameters can be expressed as,

\begin{equation}
 r=1+\frac{9}{2(\rho_{c}+\rho_{m})}\left(\frac{\partial p_{c}}{\partial
 \rho_{c}}(\rho_{c}+p_{c})+\frac{\partial p_{m}}{\partial
 \rho_{m}}(\rho_{m}+p_{m})\right)
 \end{equation}

\begin{equation}
s=\frac{1}{(p_{c}+p_{m})}\left(\frac{\partial p_{c}}{\partial
 \rho_{c}}(\rho_{c}+p_{c})+\frac{\partial p_{m}}{\partial
 \rho_{m}}(\rho_{m}+p_{m})\right)
\end{equation}

and

\begin{equation}
q=\frac{1}{2}+\frac{3}{2}\left(\frac{p_{c}+p_{m}}{\rho_{c}+\rho_{m}}\right)
\end{equation}

As a complementary to $\{r,s\}$, a new diagnostic called $Om$ has
been recently proposed, which helps to distinguish the present
matter density contrast in different models more effectively. The
new diagnostic of dark energy $Om$ is introduced to differentiate
$\Lambda$CDM from other DE models. $Om$ diagnostic is defined as
[35],
\begin{equation}
Om(z)=\frac{\left(\frac{H(z+1)}{H_0}\right)^2-1}{(z+1)^3-1}
\end{equation}
Thus $Om$ involves only the first derivative of the scale factor
through the Hubble parameter and is easier to reconstruct from
observational data. $Om$ is a constant in $\Lambda$CDM model,
since it is independent of redshift z and it provides a null test
of cosmological constant. $Om$ diagnostic can distinguish DE
models with less dependence on matter density relative to the EOS
of DE.

\section{\normalsize\bf{Different Models of Cardassian Universe}}

\subsection{Polytropic/Power Law Model (PL)}

The simplest model is the power law (PL) Cardassian model where
$g(\rho)=\rho+B\rho^n$, with $B$ and $n < 2/3$ are two constants
and the additional term $\rho^n$ satisfies many observational
constraints such as if the first Doppler peak of the CMB is
slightly shifted, the universe is rather older, and the early
structure formation $z > 1$ is unaffected. In this model $g(\rho)$
is defined as [23],
\begin{equation}
g(\rho)=\rho_{m}\left[1+\left(\frac{\rho_{m}}{\rho_{card}}\right)^{(n-1)}\right]
\end{equation}
where $\rho_{card}$ is a characteristic constant energy density
and $n$ is a dimensionless positive constants. So from (9), (14),
(16) and (22) we get,
\begin{equation}
\rho_c=\frac{\rho_0^n (1+z)^{3n(1+w_m)}}{\rho_{card}^{(n-1)}}
\end{equation}
and
\begin{equation}
p_c=\frac{(n+nw_m-1)\rho_0^n
(1+z)^{3n(1+w_m)}}{\rho_{card}^{(n-1)}}
\end{equation}
So the equation of states are given by,
\begin{equation}
w_c=\frac{p_c}{\rho_c}=(n+nw_m-1)
\end{equation}
 and
\begin{equation}
w=\frac{p}{g(\rho)}=\frac{w_m+(n+nw_m-1)\left(\frac{\rho_0
(1+z)^{3(1+w_m)}}{\rho_{card}}\right)^{(n-1)}}{1+\left(\frac{\rho_0
(1+z)^{3(1+w_m)}}{\rho_{card}}\right)^{(n-1)}}
\end{equation}

\begin{figure}[!h]
\includegraphics[height=1.8in]{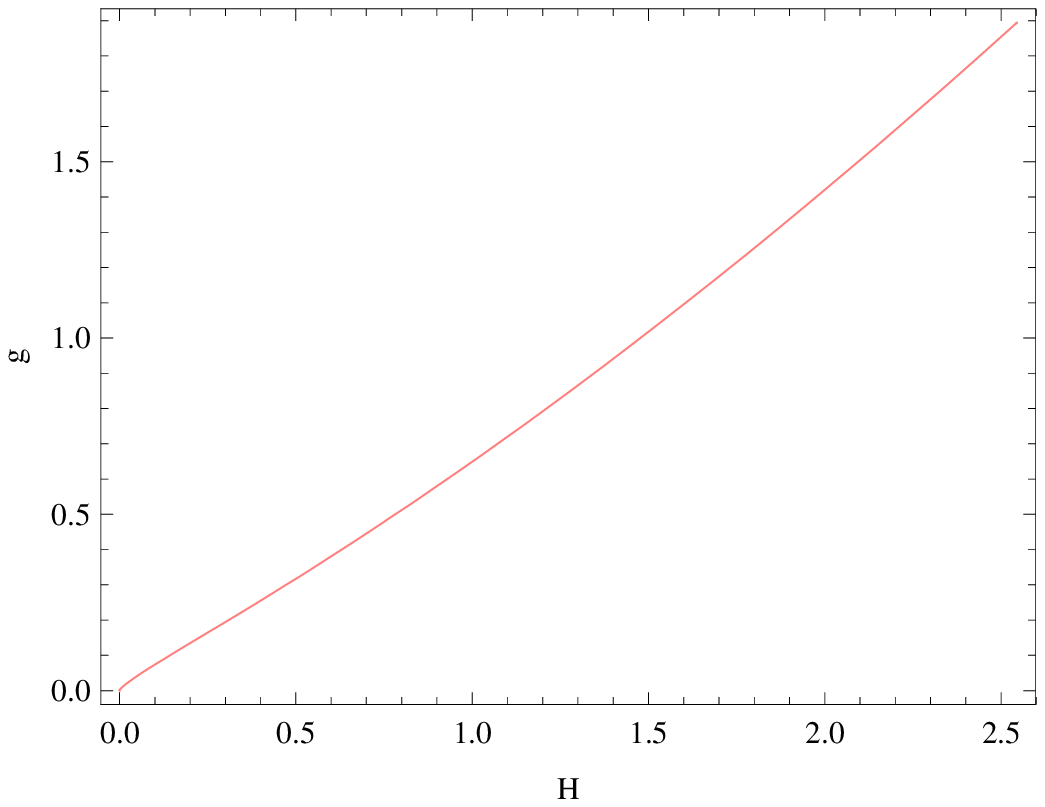}~\\
\vspace{1mm}\caption{The variation of $g$ against $H$ for $w_m =
.01, n = .5,\rho_{card} = .1,\Lambda = .01, \rho_0 = .001, k = 1,
l=.9, H_0 =72$.}
\end{figure}

\begin{figure}[!h]
\includegraphics[height=2.3in]{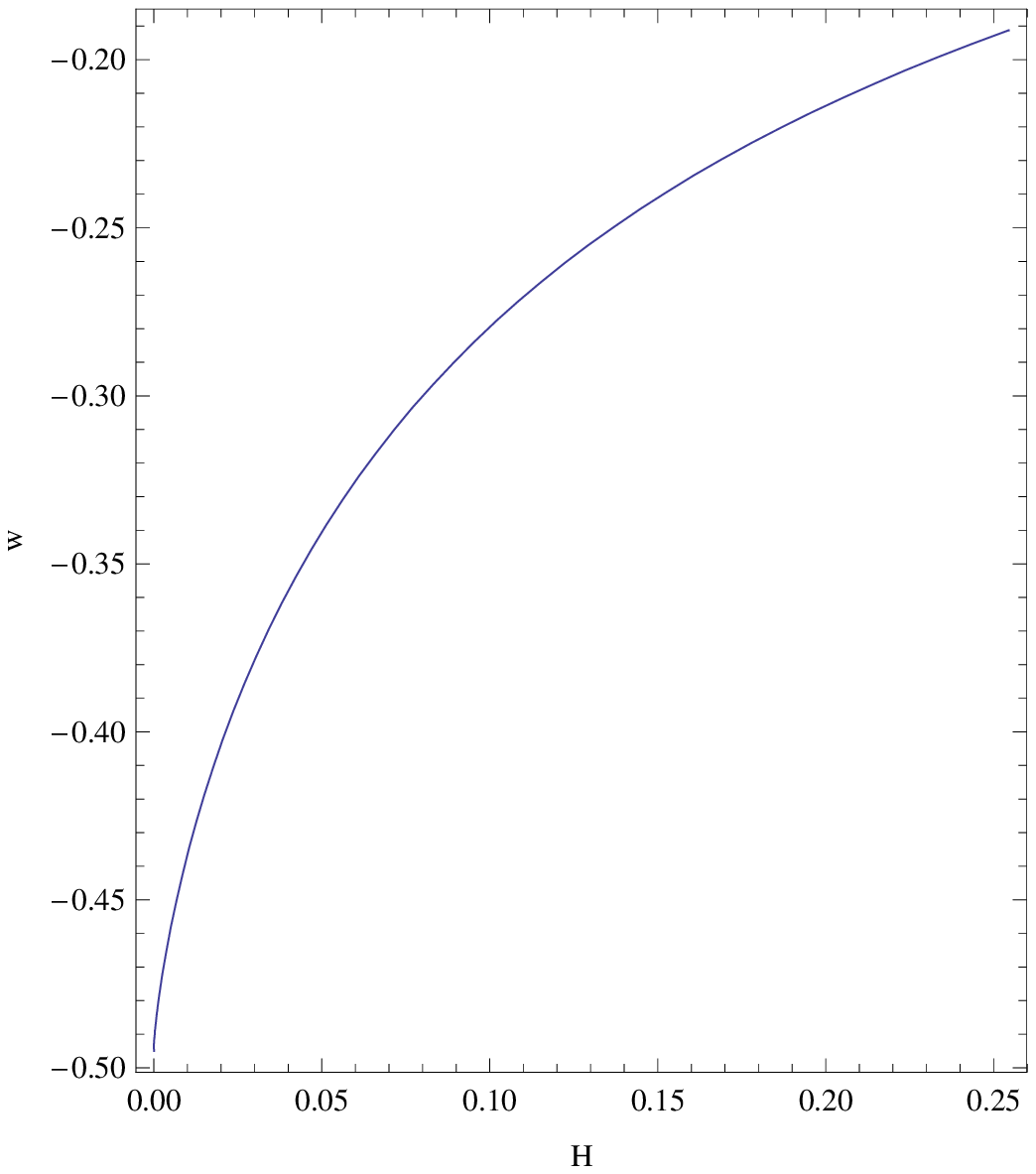}~\\
\vspace{1mm}\caption{The variation of $w$ against $H$ for $w_m =
.01, n = .5,\rho_{card} = .1,\Lambda = .01, \rho_0 = .001, k = 1,
l=.9, H_0 =72$.}
\end{figure}

\begin{figure}[!h]
\includegraphics[height=1.3in]{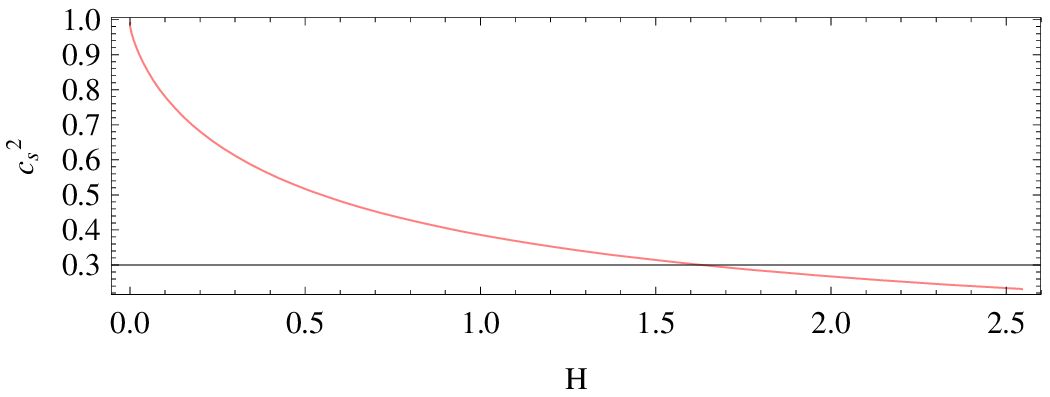}~\\
\vspace{1mm}\caption{The variation of $c_{s}^{2}$ against $H$ for
$w_m = .01, n = .5,\rho_{card} = .1,\Lambda = .01, \rho_0 = .001,
k = 1, l=.9, H_0 =72$.}
\end{figure}

\begin{figure}[!h]
\includegraphics[height=1.5in]{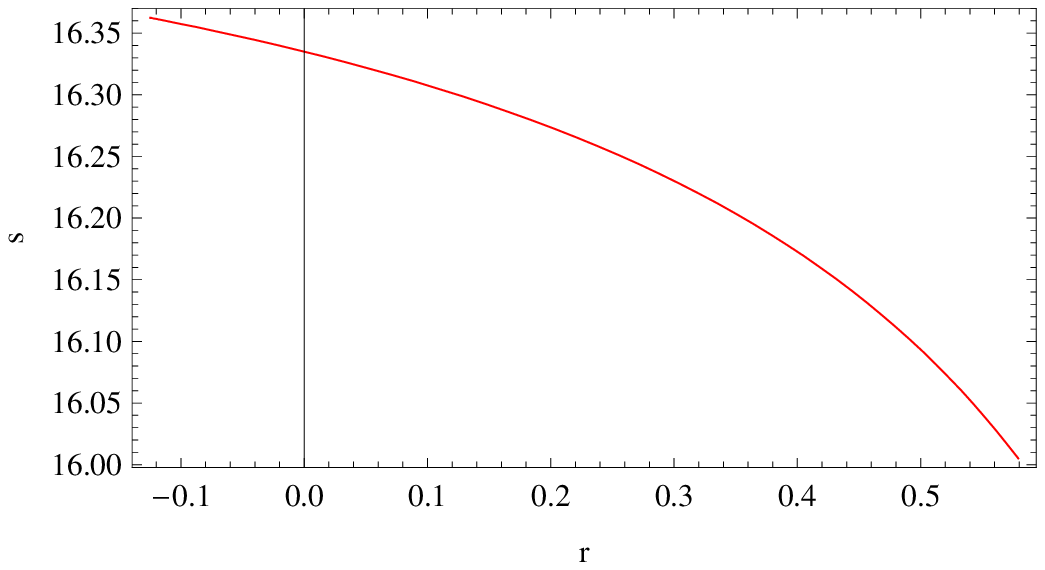}~\\
\vspace{1mm}\caption{The variation of $s$ against $r$ for $w_m =
.01, n = .5,\rho_{card} = .1,\Lambda = .01, \rho_0 = .001, k = 1,
l=.9, H_0 =72$.}
\end{figure}

\begin{figure}[!h]
\includegraphics[height=2in]{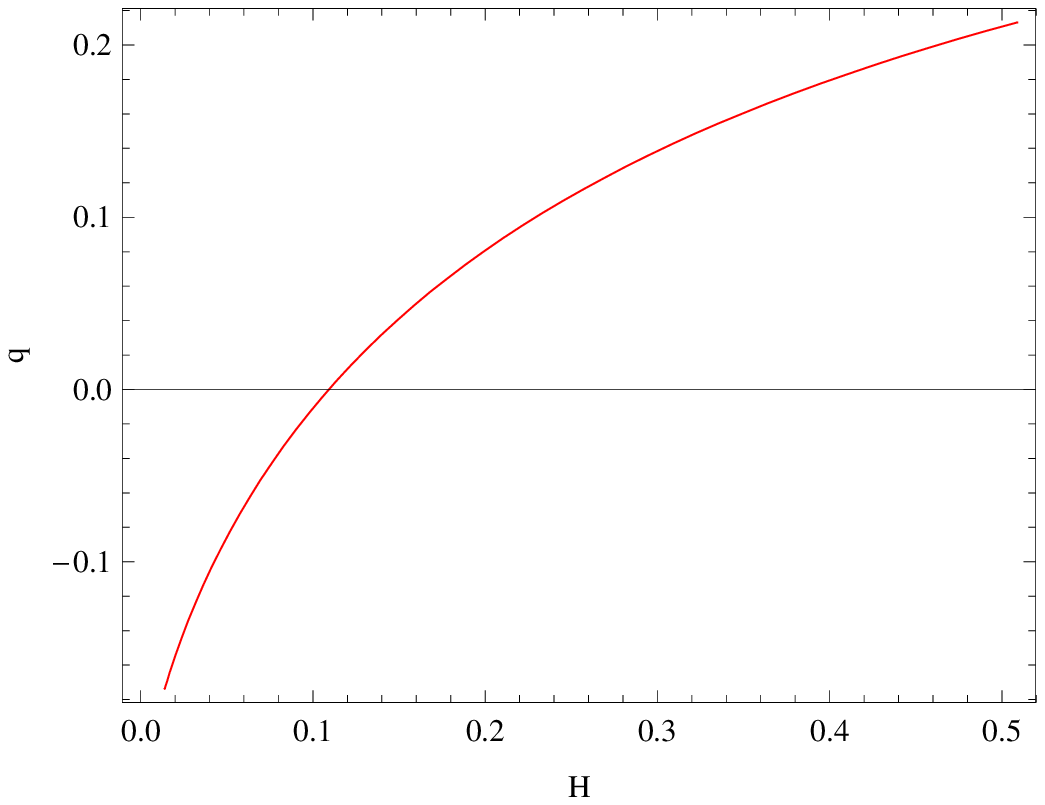}~\\
\vspace{1mm}\caption{The variation of $q$ against $H$ for $w_m =
.01, n = .5,\rho_{card} = .1,\Lambda = .01, \rho_0 = .001, k = 1,
l=.9, H_0 =72$.}
\end{figure}

\begin{figure}[!h]
\includegraphics[height=1.7in]{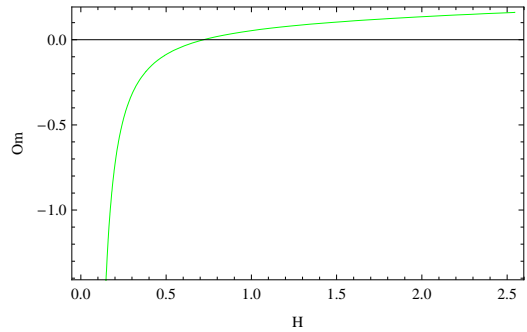}~\\
\vspace{1mm}\caption{The variation of $Om$ against $H$ for $w_m =
.01, n = .5,\rho_{card} = .1,\Lambda = .01, \rho_0 = .001, k = 1,
l=.9, H_0 =72$.}
\end{figure}

From above, we see that $w_c$ is a constant. From (16), (18),
(19), (20), (23) and (24) we get the statefinder parameters and
deceleration parameter,
\begin{equation}
r=\frac{(1 + z)^3\rho_0(2 + 9 w_m (1 + w_m)) + (1 + z)^{-3 w_m}
\left(\frac{(1 + z)^{3(1 +w_m)}\rho_0}{\rho_{card}}\right)^n
\rho_{card} (2 +9 n (1 + w_m) (-1 + n + n w_m))}{2 (1 +z)^3 \rho_0
+2 (1 + z)^{-3 w_m} \left(\frac{(1 + z)^{3 (1 + w_m)}
\rho_0}{\rho_{card}}\right)^n\rho_{card}}
\end{equation}
and
\begin{equation}
s=\frac{(1 + z)^3 \rho_0 w_m (11 + 9w_m) + (1 + z)^{-3w_m}
\left(\frac{(1 + z)^{3(1 +w_m)}\rho_0}{\rho_{card}}\right)^n
\rho_{card}(-1 + n + n w_m) (2 + 9 n (1 + w_m))}{2 (1 + z)^3
\rho_0 w_m + 2(1 + z)^{-3 w_m} \left(\frac{(1 + z)^{3(1
+w_m)}\rho_0}{\rho_{card}}\right)^n \rho_{card}(-1 + n + n w_m)}
\end{equation}

\begin{equation}
q=\frac{(1 + z)^3 \rho_0 (1 + 3 w_m) + (1 + z)^{-3w_m}
\left(\frac{(1 + z)^{3(1 +w_m)}\rho_0}{\rho_{card}}\right)^n
\rho_{card}(-2 + 3 n (1 + w_m))}{2 (1 + z)^3 \rho_0 +2(1 + z)^{-3
w_m} \left(\frac{(1 + z)^{3(1 +w_m)}\rho_0}{\rho_{card}}\right)^n
\rho_{card}}
\end{equation}

Fig.1 ~represents the variation of $g$ against $H$, Fig.2
~represents the variation of $w$ against $H$. From (9), (10),
(16), (23) and (24) we plot the graph which shows the variation of
the square of velocity of sound $c_s^2$ against  $H$ is given in
Fig.3. Fig.4 ~represents the variation of $s$ against $r$ and
Fig.5 ~represents the variation of $q$ against $H$. From (8), (21)
and (22) we get the value of $Om$. Fig.6 ~represents the variation
of $Om$ against $H$. The values are  taken as, $w_m = .01, n=.5,
\rho_{card} = .1, \Lambda = .01, \rho_0 = .001, k = 1, l=.9, H_0
=72$. From the figures, we see that $g,w,q$ and $Om$ decrease as
$H$ decreases and $c_s^2$ lies between 0 and 1. The parameter $s$
increases and keeps positive sign as $r$ decreases from positive
to negative
values.\\

\subsection{Modified Polytropic Model (MP)}

Modified polytropic Cardassian model is a slight modification of
the previous one, can be used on all scales, but it does not quite
fit the criteria of the Cardassian model as defined above. At late
times in the future of the Universe, when $\rho_m\ll\rho_{card}$,
this model becomes cosmological constant dominated with
$\Lambda=\rho_{card}$. This energy density is very similar to a
model which motivated by gravitational leakage into extra
dimensions. In this model $g(\rho)$ is defined as [25],
\begin{equation}
g(\rho)=\rho_{m}\left[1+\left(\frac{\rho_{card}}{\rho_{m}}\right)^{\alpha(n-1)}\right]^{\frac{1}{\alpha}}
\end{equation}
where $\rho_{card}$ is a characteristic constant energy density
and $\alpha\neq 0$ and $n$ are dimensionless positive constants.
So from (9), (14), (16) and (30) we get,
\begin{equation}
\rho_c=(1 + z)^{3 (1 + w_m)} \rho_0 (-1 + (1
+X_{MP})^{\frac{1}{\alpha}})
\end{equation}
and
\begin{equation}
p_c=(1 + z)^{3 (1 +w_m)}\rho_0 (X_{MP}(1 + X_{MP})^{-1 +
     \frac{1}{\alpha}} -w_m + (1 + X_{MP})^{-1 +
     \frac{1}{\alpha}}(w_m - X_{MP}(n + (-2 + n)w_m)))
\end{equation}
where, $$X_{MP}=\left[\frac{(1 +
z)^{-3(1+w_m)}\rho_{card}}{\rho_0}\right]^{(n-1)\alpha}$$

\begin{figure}[!h]
\includegraphics[height=2.1in]{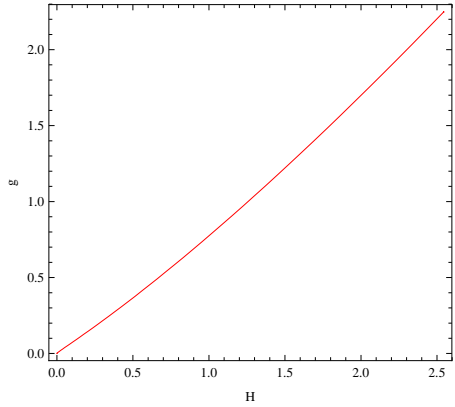}~\\
\vspace{1mm}\caption{The variation of $g$ against $H$ for
$\alpha=.5, w_m = .01, n=.5, \rho_{card} = .1,\Lambda = .01,
\rho_0 = .001, k = 1, l=.9, H_0 =72$.}
\end{figure}

\begin{figure}[!h]
\includegraphics[height=1.7in]{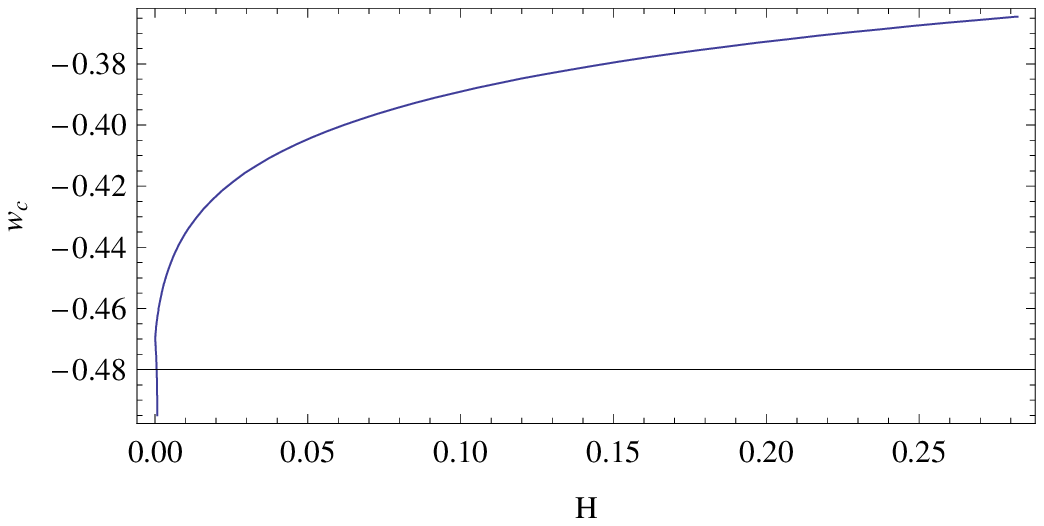}~\\
\vspace{1mm}\caption{The variation of $w_c$ against $H$ for
$\alpha=.5, w_m = .01, n=.5, \rho_{card} = .1,\Lambda = .01,
\rho_0 = .001, k = 1, l=.9, H_0 =72$.}
\end{figure}

\begin{figure}[!h]
\includegraphics[height=1.7in]{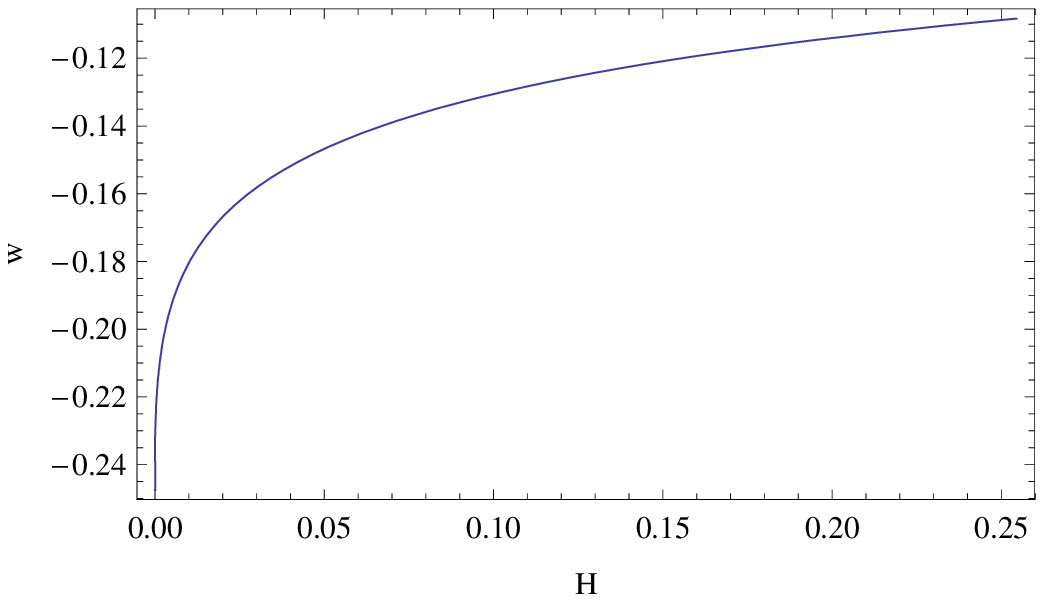}~\\
\vspace{1mm}\caption{The variation of $w$ against $H$ for
$\alpha=.5, w_m = .01, n=.5, \rho_{card} = .1,\Lambda = .01,
\rho_0 = .001, k = 1, l=.9, H_0 =72$.}
\end{figure}

\begin{figure}[!h]
\includegraphics[height=1.5in]{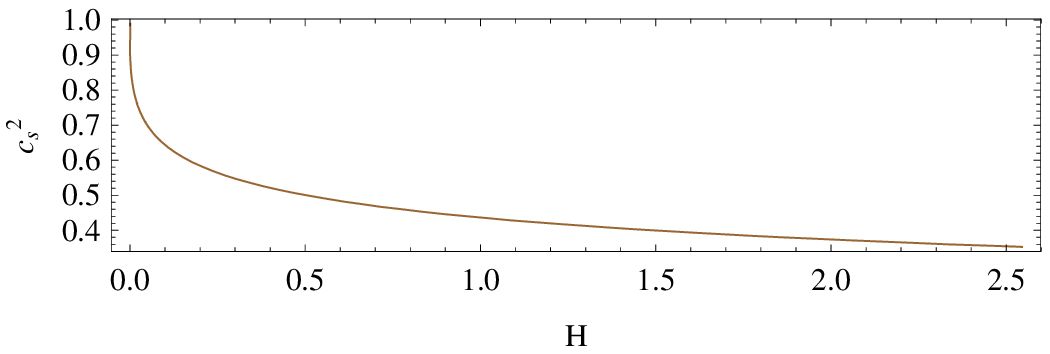}~\\
\vspace{1mm}\caption{The variation of $c_s^2$ against $H$ for
$\alpha=.5, w_m = .01, n=.5, \rho_{card} = .1,\Lambda = .01,
\rho_0 = .001, k = 1, l=.9, H_0 =72$.}
\end{figure}

\begin{figure}[!h]
\includegraphics[height=1.7in]{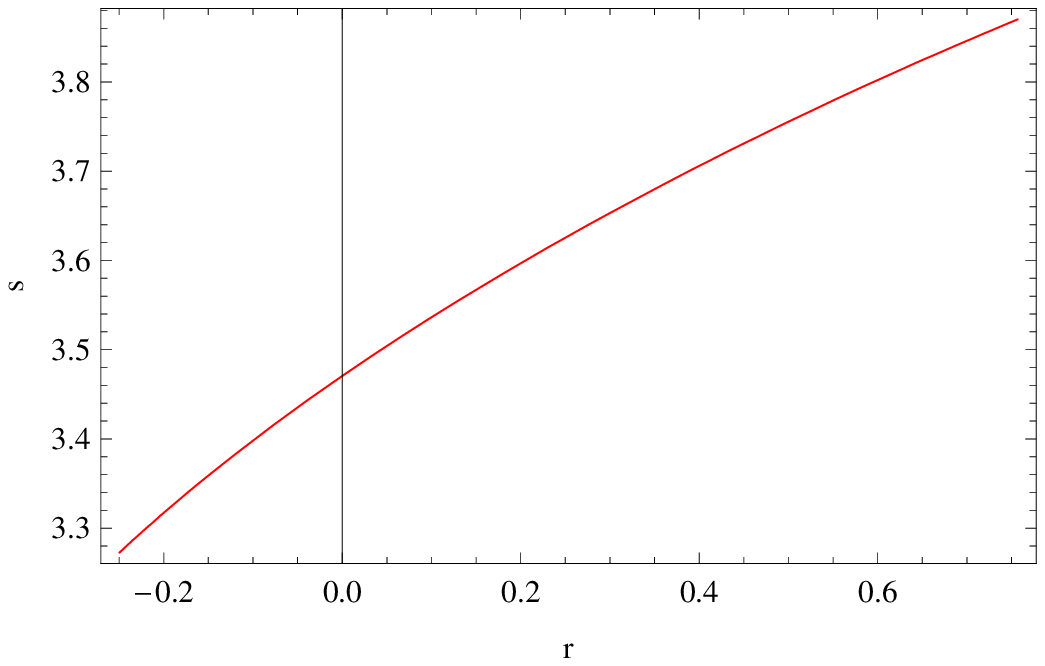}~\\
\vspace{1mm}\caption{The variation of $s$ against $r$ for
$\alpha=.5, w_m = .01, n=.5, \rho_{card} = .1,\Lambda = .01,
\rho_0 = .001, k = 1, l=.9, H_0 =72$.}
\end{figure}

\begin{figure}[!h]
\includegraphics[height=2in]{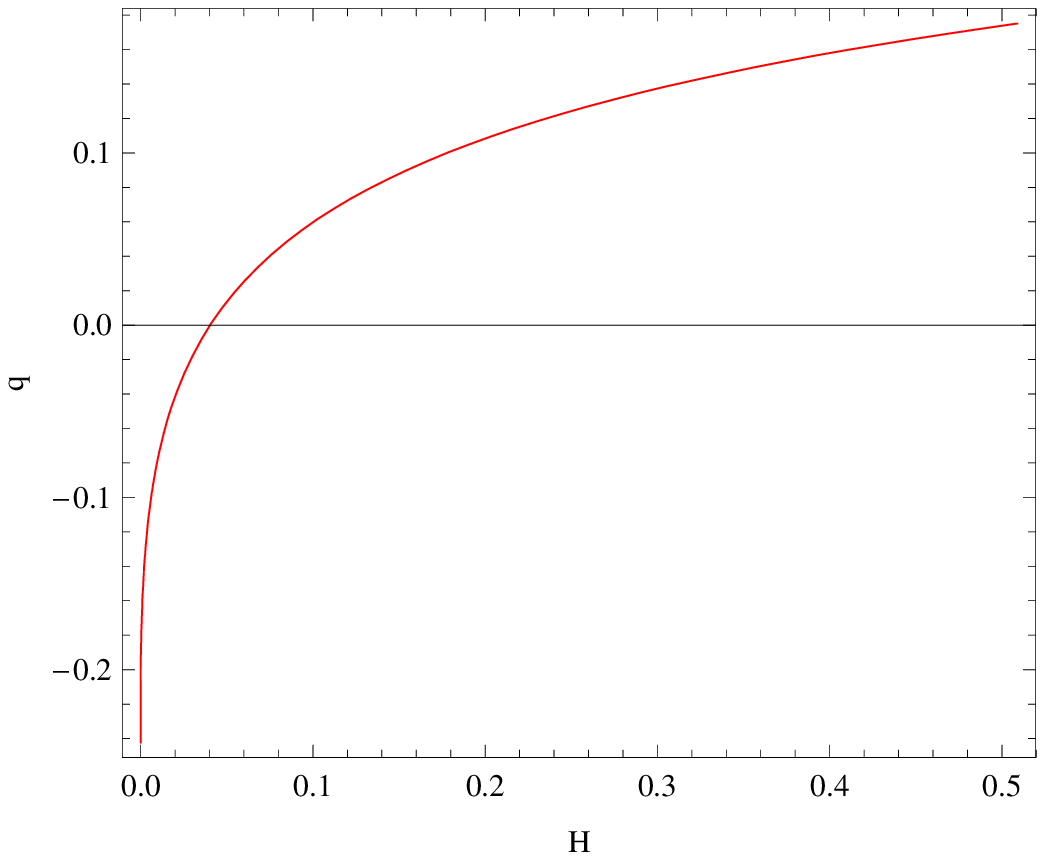}~\\
\vspace{1mm}\caption{The variation of $q$ against $H$ for
$\alpha=.5, w_m = .01, n=.5, \rho_{card} = .1,\Lambda = .01,
\rho_0 = .001, k = 1, l=.9, H_0 =72$.}
\end{figure}

\begin{figure}[!h]
\includegraphics[height=1.7in]{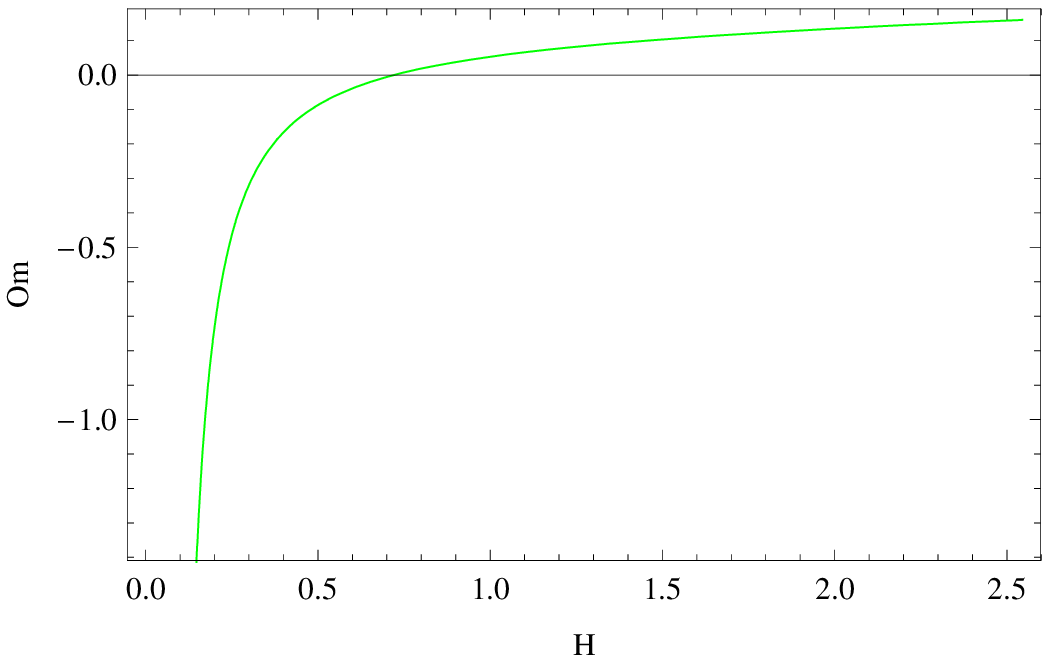}~\\
\vspace{1mm}\caption{The variation of $Om$ against $H$ for
$\alpha=.5, w_m = .01, n=.5, \rho_{card} = .1,\Lambda = .01,
\rho_0 = .001, k = 1, l=.9, H_0 =72$.}
\end{figure}

So the equation of state is given by,
\begin{equation}
w_c=\frac{(1 + z)^{3 (1 +w_m)}\rho_0 (X_{MP}(1 + X_{MP})^{-1
+\frac{1}{\alpha}} -w_m + (1 + X_{MP})^{-1 + \frac{1}{\alpha}}(w_m
- X_{MP}(n + (-2 + n)w_m)))}{(1 + z)^{3 (1 + w_m)} \rho_0 (-1 + (1
+X_{MP})^{\frac{1}{\alpha}})}
\end{equation}
 and
\begin{equation}
w=\frac{\rho_0 w_m(1+z)^{3(1+w_m)}+(1 + z)^{3 (1 +w_m)}\rho_0
(X_{MP}(1 + X_{MP})^{-1 +
     \frac{1}{\alpha}} -w_m + (1 + X_{MP})^{-1 +
     \frac{1}{\alpha}}(w_m - X_{MP}(n + (-2 + n)w_m)))}{\rho_0
(1+z)^{3(1+w_m)}+(1 + z)^{3 (1 + w_m)} \rho_0 (-1 + (1
+X_{MP})^{\frac{1}{\alpha}})}
\end{equation}

From (16), (18), (19), (20), (31) and (32) we get the statefinder
parameters and deceleration parameter,
\begin{eqnarray*}
r=\left[2 + 9w_m +9 w_m^2 + X_{MP}(-5 - 6 w_m +3 n (1 + w_m))(-4 -
6 w_m +3 n (1 + w_m))+ X_{MP}(13 +9 \alpha (1 + w_m)^2 \right.
\end{eqnarray*}
\begin{equation}
\left.+9 n^2 \alpha (1 + w_m)^2 +9 w_m (5 + 4 w_m) -9 n (1 + w_m)
(1 + 2 w_m + 2 \alpha (1 + w_m)))\right]/\left[2 (1 +
X_{MP})^2\right]
\end{equation} and

\begin{equation}
s=\frac{1}{2}\left[20 - 9 n + 18 w_m - 9 n w_m - \frac{9 (-1 + n)
(-1 + \alpha) (1 + w_m)}{1+X_{MP}} - \frac{9 (-1 + n) \alpha w_m
(1 +w_m)}{-w_m + X_{MP}(-1 + n + (-2 + n) w_m)}\right]
\end{equation}

\begin{equation}
q=\frac{1}{2}\left[4 - 3 n + 6 w_m - 3 n w_m + \frac{3 (-1 + n) (1
+ w_m)}{1 + X_{MP}}\right]
\end{equation}

Fig.7, 8 and 9 ~represents the variation of $g$, $w_c$ and $w$
against $H$ respectively. From (9), (10), (16), (31) and (32) we
plot the graph which shows the variation of the square of velocity
of sound $c_s^2$ against  $H$ is given in Fig. 10, Fig.11
~represents the variation of $s$ against $r$ and Fig.12
~represents the variation of $q$ against $H$. From (8), (21) and
(30) we get the value of $Om$. Fig.13 ~represents the variation of
$Om$ against $H$. The value of parameters are taken as,
$\alpha=.5, w_m = .01, n=.5, \rho_{card} = .1, \Lambda = .01,
\rho_0 = .001, k = 1, l=.9, H_0 =72$. From the figures, we see
that $g,w_c,w,q$ and $Om$ decrease as $H$ decreases and $c_s^2$
lies between 0 and 1. The parameter $s$ decreases and keeps
positive sign as $r$ decreases from
positive to negative values.\\

\subsection{Exponential Model:-}

In this model $g(\rho)$ is defined as [26],
\begin{equation}
g(\rho)=\rho_{m}~exp\left[\left(\frac{\rho_{card}}{\rho_{m}}\right)^n\right]
\end{equation}
where $\rho_{card}$ is a characteristic constant energy density
and $n$ is a dimensionless positive constants. So from (9), (14),
(16) and (38) we get,
\begin{equation}
\rho_c=(-1 + e^{X_{Exp}})(1 + z)^{3 (1 + w_m)} \rho_0
\end{equation}
and
\begin{equation}
p_c=-(1 + z)^{3 (1 + w_m)} \rho_0 (w_m + e^{X_{Exp}} (-w_m +n
X_{Exp}(1 + w_m)))
\end{equation}
where,
$$X_{Exp}=\left[\frac{(1+z)^{-3(1+w_m)}\rho_{card}}{\rho_0}\right]^n$$
So the equation of state is given by,
\begin{equation}
w_c=\frac{-(1 + z)^{3 (1 + w_m)} \rho_0 (w_m + e^{X_{Exp}} (-w_m
+n X_{Exp}(1 + w_m)))}{(-1 + e^{X_{Exp}})(1 + z)^{3 (1 + w_m)}
\rho_0}
\end{equation}
 and
\begin{equation}
w=\frac{\rho_0w_0(1 + z)^{3 (1 + w_m)}-(1 + z)^{3 (1 + w_m)}
\rho_0 (w_m + e^{X_{Exp}}(-w_m +n X_{Exp}(1 + w_m)))}{\rho_0(1 +
z)^{3 (1 + w_m)}+(-1 + e^{X_{Exp}})(1 + z)^{3 (1 + w_m)} \rho_0}
\end{equation}

From (16), (18), (19), (20), (39) and (40) we get the statefinder
parameters and deceleration parameter,
\begin{equation}
r=\frac{1}{2}\left(2 + 9 w_m + 9 (w_m^2 +n^2 X_{Exp}^2(1 + w_m)^2
+n X_{Exp}(1 + w_m) (-1 + n + (-2 + n) w_m))\right)
\end{equation}
and
\begin{equation}
s=\frac{1}{2}\left(11 - 9 n + 9 w_m - 9 n w_m -9 n X_{Exp}(1 +
w_m) - \frac{9 n w_m (1 + w_m)}{-w_m + n X_{Exp}(1 + w_m)}\right)
\end{equation}

\begin{equation}
q=\frac{1}{2}\left[1 + 3 w_m -3 n X_{Exp}(1 + w_m)\right]
\end{equation}

\begin{figure}[!h]
\includegraphics[height=2.1in]{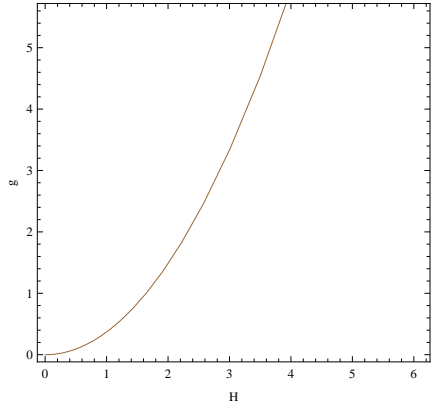}~\\
\vspace{1mm}\caption{The variation of $w$ against $H$ for $w_m =
.01, n=2, \rho_{card} = .1,\Lambda = .01, \rho_0 = .001, k = 1,
l=.9, H_0 =72$.}
\end{figure}

\begin{figure}[!h]
\includegraphics[height=1.7in]{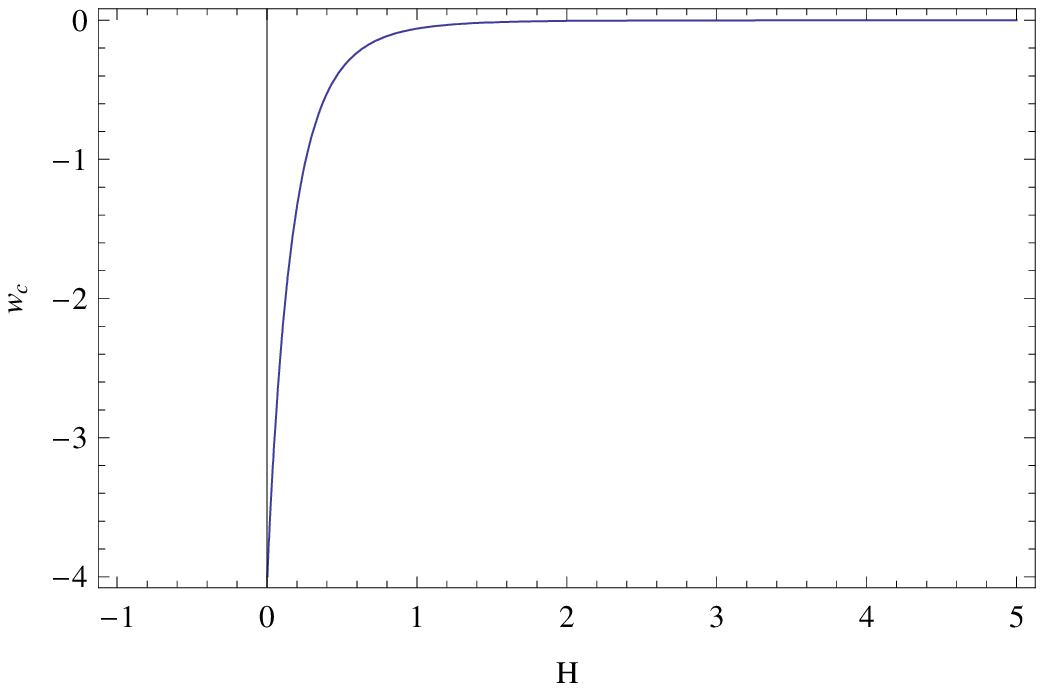}~\\
\vspace{1mm}\caption{The variation of $w$ against $z$ for $w_m =
.01, n=2, \rho_{card} = .1,\Lambda = .01, \rho_0 = .001, k = 1,
l=.9, H_0 =72$.}
\end{figure}

\begin{figure}[!h]
\includegraphics[height=1.7in]{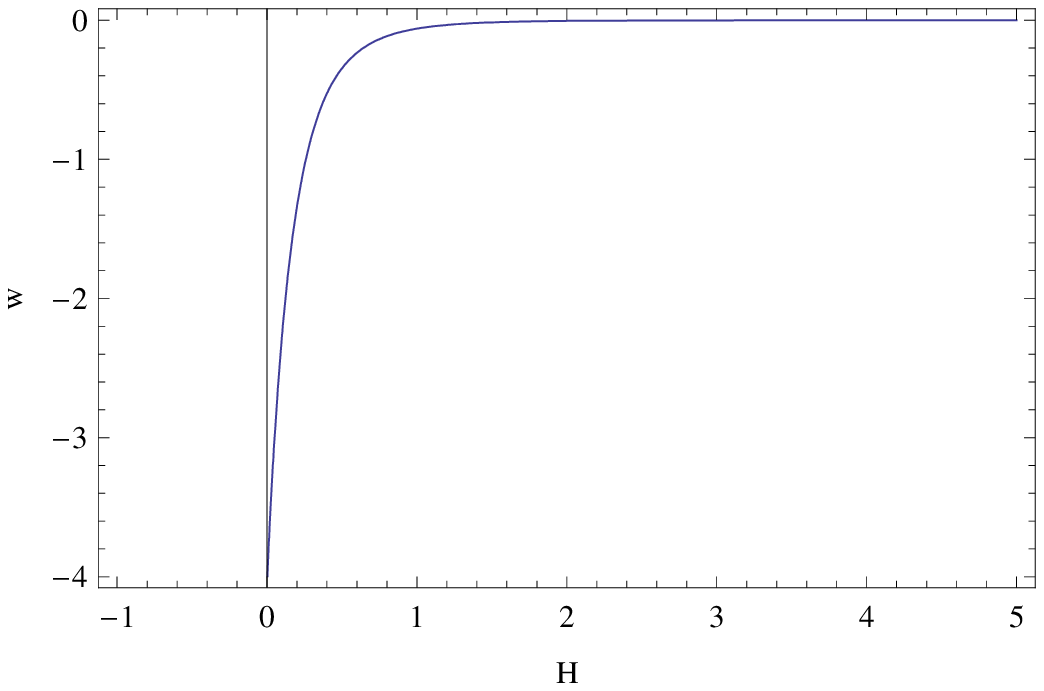}~\\
\vspace{1mm}\caption{The variation of $w$ against $z$ for $w_m =
.01, n=2, \rho_{card} = .1,\Lambda = .01, \rho_0 = .001, k = 1,
l=.9, H_0 =72$.}
\end{figure}

\begin{figure}[!h]
\includegraphics[height=1.7in]{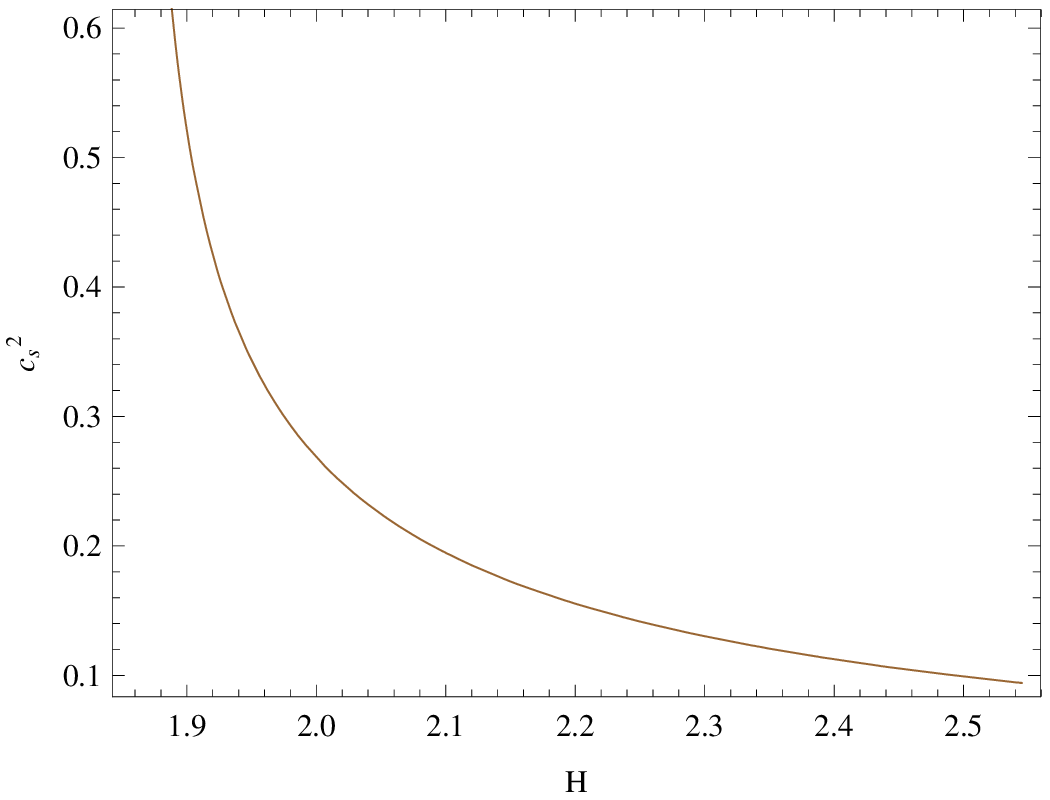}~\\
\vspace{1mm}\caption{The variation of $c_s^2$ against $H$ for $w_m
= .01, n=2, \rho_{card} = .1,\Lambda = .01, \rho_0 = .001, k = 1,
l=.9, H_0 =72$.}
\end{figure}

\begin{figure}[!h]
\includegraphics[height=1.7in]{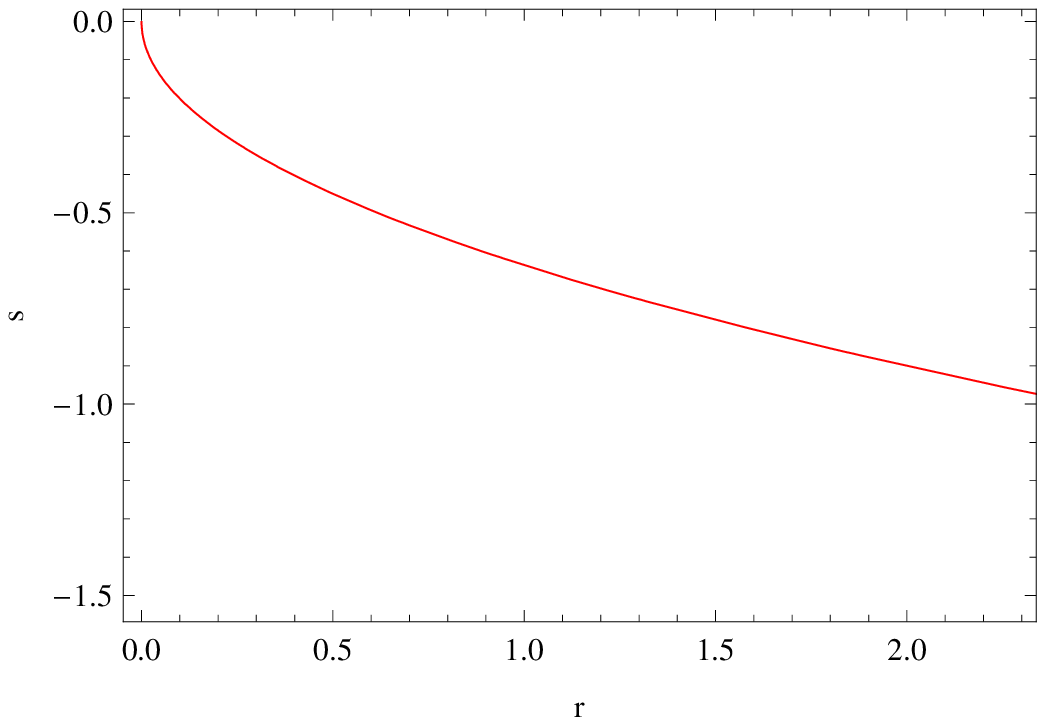}~\\
\vspace{1mm}\caption{The variation of $s$ against $r$ for $w_m =
.01, n=2, \rho_{card} = .1,\Lambda = .01, \rho_0 = .001, k = 1,
l=.9, H_0 =72$.}
\end{figure}

\begin{figure}[!h]
\includegraphics[height=1.7in]{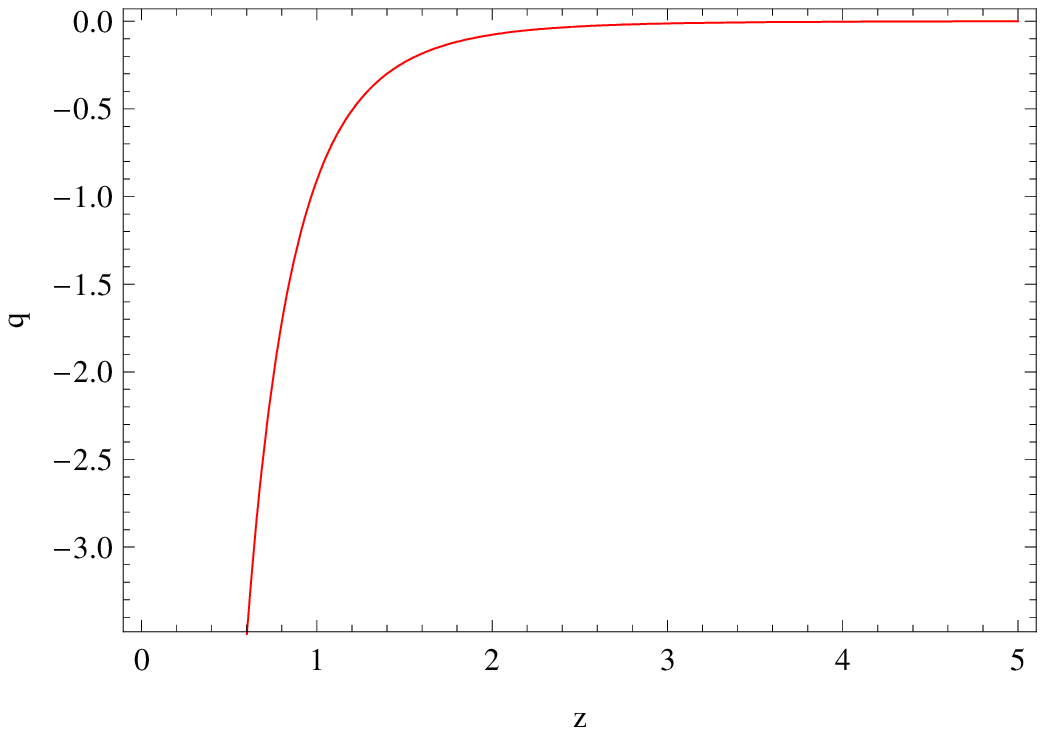}~\\
\vspace{1mm}\caption{The variation of $q$ against $z$ for $w_m =
.01, n=2, \rho_{card} = .1,\Lambda = .01, \rho_0 = .001, k = 1,
l=.9, H_0 =72$.}
\end{figure}

\begin{figure}[!h]
\includegraphics[height=1.7in]{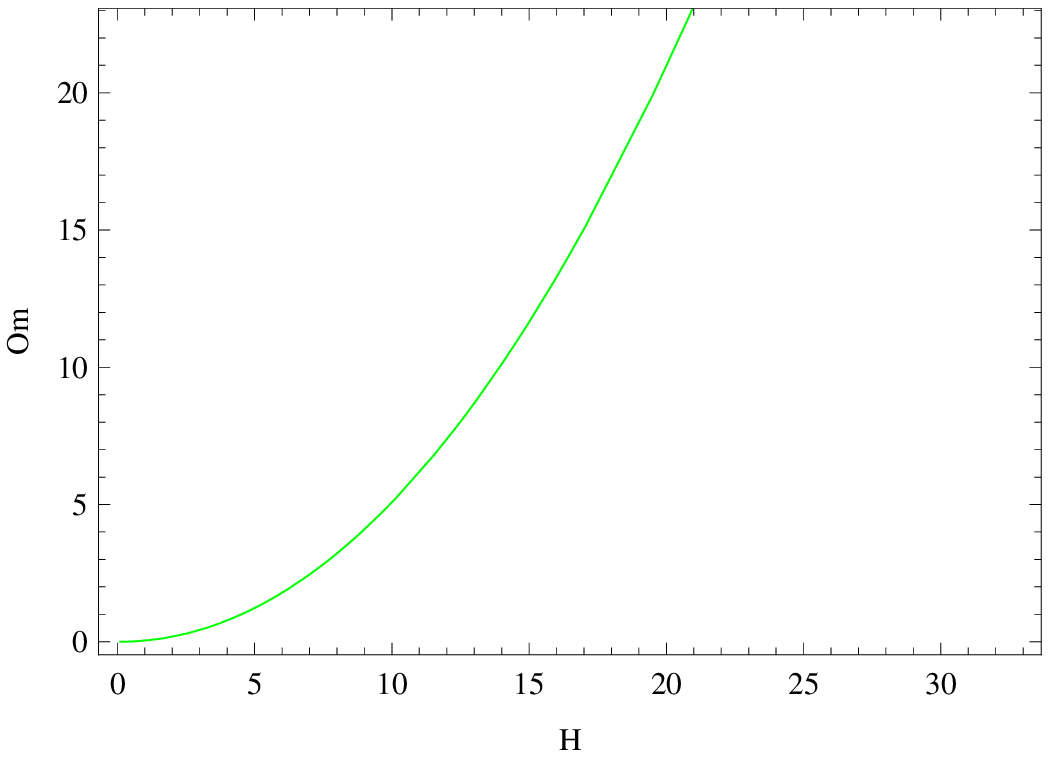}~\\
\vspace{1mm}\caption{The variation of $Om$ against $H$ for $w_m =
.01, n=2, \rho_{card} = .1,\Lambda = .01, \rho_0 = .001, k = 1,
l=.9, H_0 =72$.}
\end{figure}

Fig.14, 15 and 16 ~represents the variation of $g$, $w_c$ and $w$
against $H$, $z$ and $z$ respectively. From (9), (10), (16), (39)
and (40) we plot the graph which shows the variation of the square
of velocity of sound $c_s^2$ against  $H$ is given in Fig. 17.
Fig.18 ~represents the variation of $s$ against $r$ and Fig.19
~represents the variation of $q$ against $H$. From (8), (21) and
(38) we get the value of Om. Fig.20 ~represents the variation of
$Om$ against $H$. The values are taken as, $w_m = .01, n=2,
\rho_{card} = .1, \Lambda = .01, \rho_0 = .001, k = 1, l=.9, H_0
=72$. From the figures, we see that $g,w_c,w,q$ and $Om$ decrease
as $H$ decreases and $c_s^2$ lies between 0 and 1. The parameter
$s$ increases and keeps negative sign as $r$ decreases.\\

\subsection{Modified Exponential Model}

In this model $g(\rho)$ is defined as [26],
\begin{equation}
g(\rho)=(\rho_{m}+\rho_{card})~exp\left[\left(\frac{\alpha\rho_{card}}{\rho_m+\rho_{card}}\right)^n\right]
\end{equation}
where $\rho_{card}$ is a characteristic constant energy density
and $\alpha$ and $n$ are dimensionless positive constants. So from
(9), (14), (16) and (46) we get,
\begin{equation}
\rho_c=-(1 +z)^{3 (1 + w_m)} \rho_0 + e^{\left(\frac{\alpha
\rho_{card}}{(1 + z)^{3 (1 + w_m)}\rho_0 + \rho_{card}}\right)^n}
((1 + z)^{3 (1 + w_m)} \rho_0 + \rho_{card})
\end{equation}
and
\begin{equation}
p_c=(1 + z)^{3 w_m} (-(1 +z)^3 \rho_0 w_m -e^{X_{ME}}((1 + z)^{-3
w_m} \rho_{card} + (1 + z)^3 \rho_0 (-w_m +n X_{ME})^n (1 +
w_m))))
\end{equation}
So the equation of state is given by,
\begin{equation}
w_c=\frac{(1 + z)^{3 w_m} (-(1 +z)^3 \rho_0 w_m -e^{X_{ME}}((1 +
z)^{-3 w_m} \rho_{card} + (1 + z)^3 \rho_0 (-w_m +nX_{ME}(1 +
w_m))))}{-(1 +z)^{3 (1 + w_m)} \rho_0 + e^{X_{ME}} ((1 + z)^{3 (1
+ w_m)} \rho_0 + \rho_{card})}
\end{equation}
 and
\begin{equation}
w=\frac{w_m\rho_{0}(1+z)^{3(1+w_m)}+(1 + z)^{3 w_m} (-(1 +z)^3
\rho_0 w_m -e^{X_{ME}}((1 + z)^{-3 w_m} \rho_{card} + (1 + z)^3
\rho_0 (-w_m +nX_{ME}(1 + w_m))))}{\rho_{0}(1+z)^{3(1+w_m)}-(1
+z)^{3 (1 + w_m)} \rho_0 + e^{X_{ME}} ((1 + z)^{3 (1 + w_m)}
\rho_0 + \rho_{card})}
\end{equation}
where $$X_{ME}=\left(\frac{\alpha \rho_{card}}{(1 + z)^{3 (1 +
w_m)} \rho_0 + \rho_{card}}\right)^n$$

\begin{figure}
\includegraphics[height=1.7in]{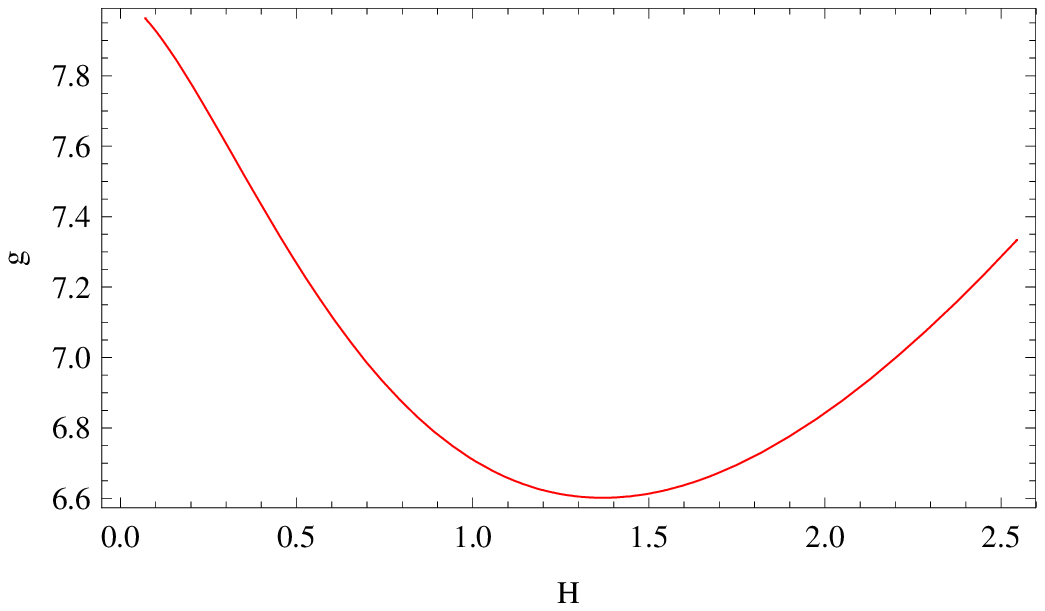}~\\
\vspace{1mm}\caption{The variation of $w_c$ against $H$ for
$\alpha=2.5, w_m = .01, n=.8, \rho_{card} = .1,\Lambda = .01,
\rho_0 = .001, k = 1, l=.9, H_0 =72$.}
\end{figure}

\begin{figure}
\includegraphics[height=1.4in]{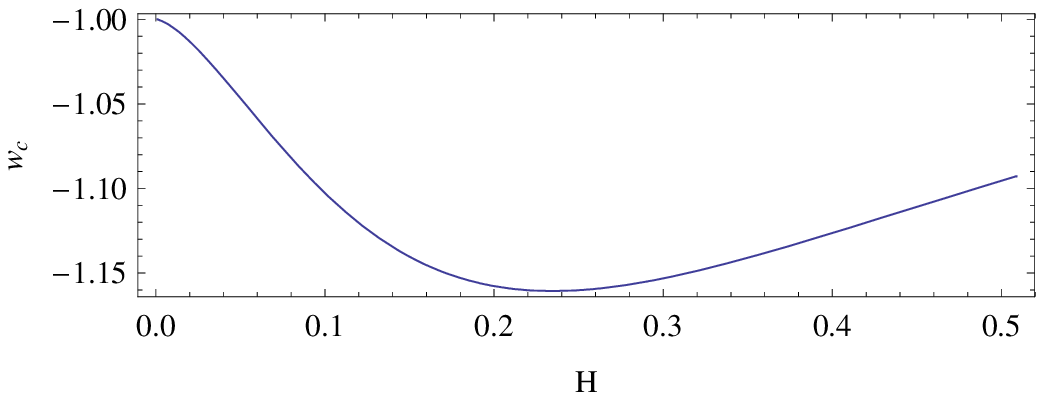}~\\
\vspace{1mm}\caption{The variation of $w_c$ against $H$ for
$\alpha=2.5, w_m = .01, n=.8, \rho_{card} = .1,\Lambda = .01,
\rho_0 = .001, k = 1, l=.9, H_0 =72$.}
\end{figure}

\begin{figure}
\includegraphics[height=1.7in]{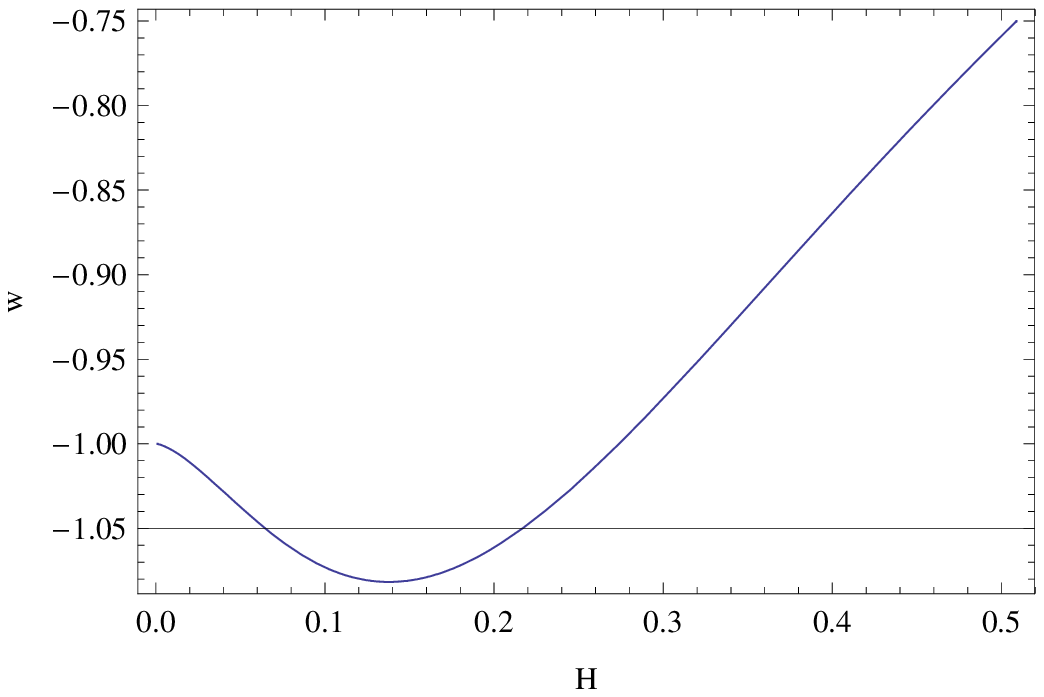}~\\
\vspace{1mm}\caption{The variation of $w$ against $H$ for
$\alpha=2.5, w_m = .01, n=.8, \rho_{card} = .1,\Lambda = .01,
\rho_0 = .001, k = 1, l=.9, H_0 =72$.}
\end{figure}

\begin{figure}
\includegraphics[height=1.7in]{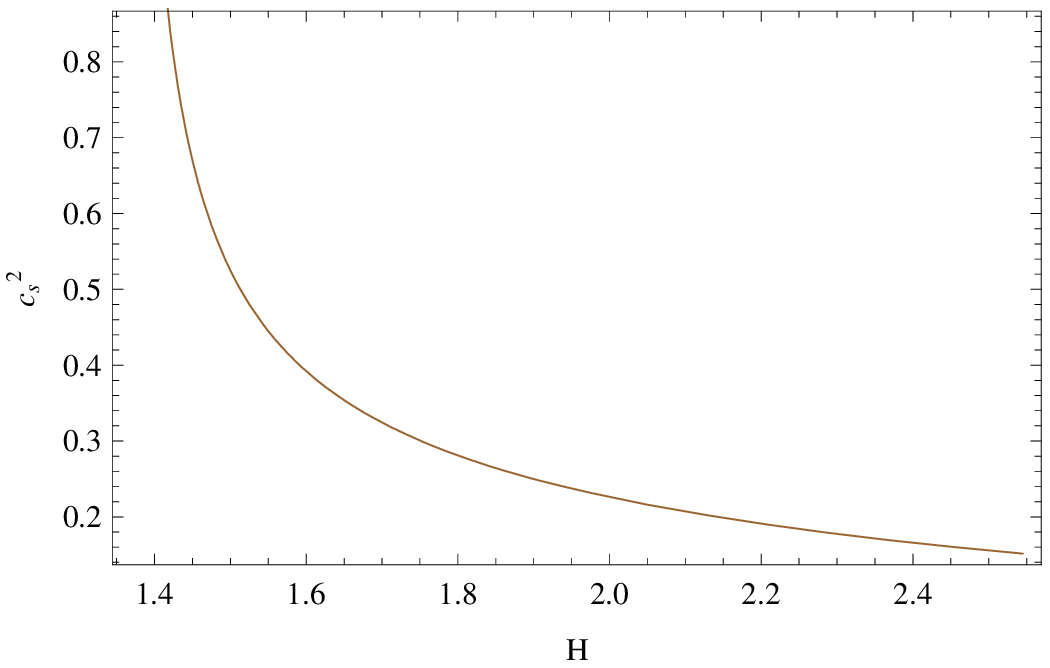}~\\
\vspace{1mm}\caption{The variation of $c_s^2$ against $H$ for
$\alpha=2.5, w_m = .01, n=.8, \rho_{card} = .1,\Lambda = .01,
\rho_0 = .001, k = 1, l=.9, H_0 =72$.}
\end{figure}

\begin{figure}
\includegraphics[height=1.7in]{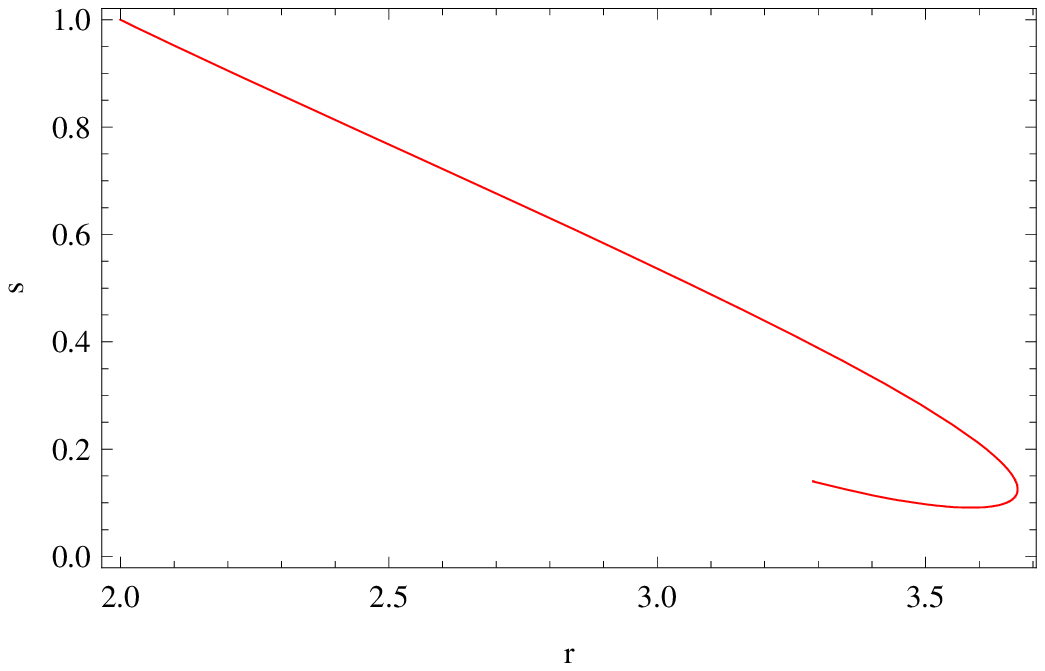}~\\
\vspace{1mm}\caption{The variation of $s$ against $r$ for
$\alpha=2.5, w_m = .01, n=.8, \rho_{card} = .1,\Lambda = .01,
\rho_0 = .001, k = 1, l=.9, H_0 =72$.}
\end{figure}

\begin{figure}
\includegraphics[height=1.2in]{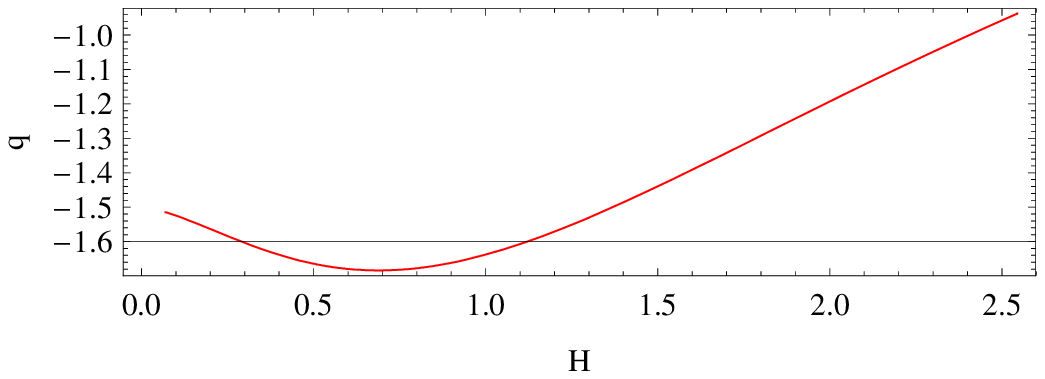}~\\
\vspace{1mm}\caption{The variation of $q$ against $H$ for
$\alpha=2.5, w_m = .01, n=.8, \rho_{card} = .1,\Lambda = .01,
\rho_0 = .001, k = 1, l=.9, H_0 =72$.}
\end{figure}

\begin{figure}
\includegraphics[height=1.7in]{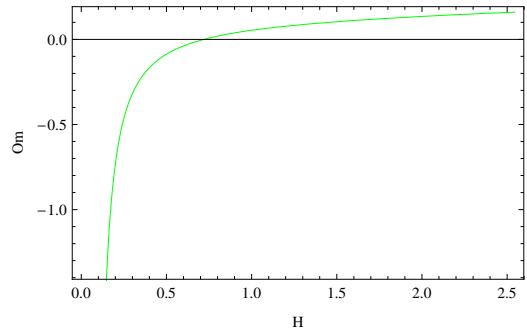}~\\
\vspace{1mm}\caption{The variation of $Om$ against $H$ for
$\alpha=2.5, w_m = .01, n=.8, \rho_{card} = .1,\Lambda = .01,
\rho_0 = .001, k = 1, l=.9, H_0 =72$.}
\end{figure}

From (16), (18), (19), (20), (47) and (48) we get the statefinder
parameters and deceleration parameter,
\begin{eqnarray*}
r=\left[2(1 + z)^{-6 w_m} \rho_{card}^2 + (1 + z)^{3-3 w_m} \rho_0
\rho_{card}(4 - 9 (-1 + n X_{ME}) w_m (1 + w_m))+ (1 + z)^6
\rho_0^2 (2 + 9 w_m \right.
\end{eqnarray*}
\begin{equation}
\left.+9 (w_m^2 +n^2 X_{ME}^2(1 + w_m)^2 +n X_{ME}(1 + w_m) (-1 +
n + (-2 + n) w_m)))\right]/\left[2(1 + z)^{-3 w_m}
\rho_{card}^2+2(1 +z)^3 \rho_0\right]
\end{equation}
and
\begin{eqnarray*}
s=\left[2(1 + z)^{-6 w_m} \rho_{card}^2 +(1 + z)^{3-3 w_m} \rho_0
\rho_{card} (2 + n X_{ME}(1 + w_m) (2 + 9 w_m) - w_m (11 +9
w_m))\right.
\end{eqnarray*}
\begin{eqnarray*}
\left.- (1 +z)^6 \rho_0^2 (9 n^2 X_{ME}^2(1 + w_m)^2 + w_m (11 + 9
w_m) + n X_{ME}(1 + w_m) (-11 + 9 n + 9 (-2 + n) w_m))\right]
\end{eqnarray*}
\begin{equation}
/\left[(2 ((1 +z)^3 \rho_0 + (1 + z)^{-3 w_m} \rho_{card}) ((1 +
z)^{-3 w_m} \rho_{card} + (1 + z)^3 \rho_0 (-w_m +n X_{ME}(1 +
w_m))))\right]
\end{equation}

\begin{equation}
q=-\frac{2 (1 + z)^{-3 w_m} \rho_{card} + (1 + z)^3 \rho_0 (-1 - 3
w_m +3nX_{ME}(1 + w_m))}{2(1 + z)^3 \rho_0 + 2(1 + z)^{-3 w_m}
\rho_{card}}
\end{equation}

Fig.21, 22 and 23 ~represents the variation of $g$, $w_c$ and $w$
against $H$ respectively. From (9), (10), (16), (47) and (48) we
plot the graph which shows the variation of the square of velocity
of sound $c_s^2$ against  $H$ is given in Fig. 24, Fig.25
~represents the variation of $s$ against $r$ and Fig.26
~represents the variation of $q$ against $H$. From (8), (21) and
(46) we get the value of Om. Fig.27 ~represents the variation of
$Om$ against $H$. The values are taken as, $\alpha=2.5, w_m = .01,
n=.8, \rho_{card} = .1,\Lambda = .01, \rho_0 = .001, k = 1, l=.9,
H_0 =72$. From the figures, we see that $g,w_c,w,q$ first decrease
then increase and $Om$ decreases as $H$ decreases and $c_s^2$ lies
between 0 and 1. The parameter $s$ increases and keeps positive
sign as $r$
decreases.\\

\section{\normalsize\bf{Discussions}}

In this work, the we have considered Cardassian Universe in
Ho$\check{\text r}$ava-Lifshitz gravity. The energy density and
pressure for Cardassian term have been found. Four types of
Cardassian Universe models i.e., polytropic/power law, modified
polytropic, exponential and modified exponential models have been
considered for accelerating models. To investigate the natures of
statefinder parameters, deceleration parameter, $Om$ diagnostic
and EoS parameters for all types of Cardassian models in
Ho$\check{\text r}$ava-Lifshitz gravity, we have drawn all parameters
w.r.t. Hubble parameters $H$.\\

In polytropic/power law Cardassian model of the universe $g,w,q$
and $Om$ parameters have been drawn in figures 1, 2, 5 and 6. We
have seen that $g,w,q$ and $Om$ decrease as $H$ decreases. The
values are taken as, $w_m = .01, n=.5, \rho_{card} = .1, \Lambda =
.01, \rho_0 = .001, k = 1, l=.9, H_0 =72$. In this model, the EoS
parameter $w_c$ is constant. The variation of the square of
velocity of sound $c_s^2$ against $H$ is given in fig.3 and it has
been observed that $c_s^2$ lies between 0 and 1. Also fig.4
represents the variation of $s$ against $r$ and the figure shows
that $s$ increases and keeps positive sign as $r$ decreases
from positive to negative values.\\

In modified polytropic Cardassian model of the universe
$g,w_c,w,q$ and $Om$ parameters have been drawn in figures 7, 8,
9, 12 and 13. We have seen that $g,w_c,w,q$ and $Om$ decrease as
$H$ decreases. The values are taken as, $\alpha=.5, w_m = .01,
n=.5, \rho_{card} = .1, \Lambda = .01, \rho_0 = .001, k = 1, l=.9,
H_0 =72$. The variation of the square of velocity of sound $c_s^2$
against $H$ is given in fig.10 and it has been observed that
$c_s^2$ lies between 0 and 1. Also fig.11 represents the variation
of $s$ against $r$ and the figure shows that $s$ increases and
keeps positive sign as $r$ decreases from positive to negative values.\\

In exponential Cardassian model of the universe $g,w_c,w,q$ and
$Om$ parameters have been drawn in figures 14, 15, 16, 19 and 20.
We have seen that $g,w_c,w,q$ and $Om$ decrease as $H$ decreases.
The values are taken as, $w_m = .01, n=2, \rho_{card} = .1,
\Lambda = .01, \rho_0 = .001, k = 1, l=.9, H_0 =72$. The variation
of the square of velocity of sound $c_s^2$ against $H$ is given in
fig.17 and it has been observed that $c_s^2$ lies between 0 and 1.
Also fig.18 represents the variation of $s$ against $r$ and the
figure shows that $s$ increases and keeps negative sign as $r$ decreases.\\

In modified exponential Cardassian model of the universe
$g,w_c,w,q$ and $Om$ parameters have been drawn in figures 21, 22,
23, 26 and 27. We have seen that $g,w_c,w,q$ first decrease then
increase and $Om$ decreases as $H$ decreases. The values are taken
as, $\alpha=2.5, w_m = .01, n=.8, \rho_{card} = .1,\Lambda = .01,
\rho_0 = .001, k = 1, l=.9, H_0 =72$. The variation of the square
of velocity of sound $c_s^2$ against $H$ is given in fig.24 and it
has been observed that $c_s^2$ lies between 0 and 1. Also fig.25
represents the variation of $s$ against $r$ and the
figure shows that $s$ increases and keeps positive sign as $r$ decreases.\\

{\bf Acknowledgement:}\\

The authors are thankful to IUCAA, Pune, India for warm
hospitality where part of the work was carried out.\\

{\bf References:}\\\\
$[1]$ A. Riess, et al., Astron. J. 116 (1998) 1009.\\
$[2]$ S. J. Perlmutter, et al., Astrophys. J. 517 (1999) 565.\\
$[3]$ J. L. Tonry, et al., Astrophys. J. 594 (2003) 1.\\
$[4]$ B. Barris, et al., Astrophys. J. 602 (2004) 571.\\
$[5]$ R. Knop, et al., Astrophys. J. 598 (2003) 102.\\
$[6]$ A.G. Riess, et al., Astrophys. J. 607 (2004) 665.\\
$[7]$ P. Astier, et al., Astron. Astrophys. 447 (2006) 31.\\
$[8]$ P. J. Steinhardt, L. Wang and I. Zlater, Phys. Rev. D
59(1999) 123504. \\
$[9]$ X. Z. Li, J. G. Hao and D. J. Liu, Class. Quant. Grav. 19
(2002) 6049.\\
$[10]$ D. J. Liu and X. Z. Li, Phys. Lett. B 611 (2005) 8.\\
$[11]$ K. Freese, F.C. Adams, J.A. Frieman, and E. Mottola, Nucl.
Phys. B287, 797(1987).\\
$[12]$ P.J.E. Peebles and B. Ratra, Astrophys. J., Lett. Ed. 325,
L17(1988).\\
$[13]$ B. Ratra and P.J.E. Peebles, Phys. Rev. D 37, 3406(1988).\\
$[14]$ J. Frieman, C. Hill, A. Stebbins, and I. Waga, Phys. Rev.
Lett. 75, 2077(1995).\\
$[15]$ L. Wang and P. Steinhardt, Astrophys. J. 508, 483(1998).\\
$[16]$ R. Caldwell, R. Dave, and P. Steinhardt, Phys. Rev. Lett.
80, 1582(1998).\\
$[17]$ G. Huey, L. Wang, R. Dave, R. Caldwell, and P. Steinhardt,
Phys. Rev. D 59, 063005(1999).\\
$[18]$ C. Deffayet, Phys. Lett. B 502, 199(2001).\\
$[19]$ C. Deffayet, G. Dvali, and G. Gabadadze, Phys. Rev. D 65,
044023(2002).\\
$[20]$ C.B. Netterfield et al, Astrophys. J. 571 (2002) 604.\\
$[21]$ R. Stompor et al, Astrophys. J. 561 (2001) L7.\\
$[22]$ K. Freese, M. Lewis, Phys. Lett. B 540 (2002)1.\\
$[23]$ K. Freese, New Astron. Rev. 49 (2005) 103; Nucl. Phys. (Proc. Suppl.) 124 50 (2003).\\
$[24]$ R. Lazkoz, G. León, Phys. Rev. D 71 (2005) 123516.\\
$[25]$ Y. Wang, K. Freese, P. Gondolo, M. Lewis, Astrophys. J.
594(2003) 25.\\
$[26]$ D. J. Liu, C. B. Sun and X. Z. Li, Phys. Lett. B 634, 442
(2006).\\
$[27]$ C. Pryke et al, Astrophys. J. 568, 46 (2002); N.W. Halverson et al, Astrophys. J. 568 (2002) 38.\\
$[28]$ P. Ho$\check{\text r}$ava, JHEP 0903 020 (2009). \\
$[29]$ P. Ho$\check{\text r}$ava, Phys. Rev. D 79 084008 (2009).\\
$[30]$ E. M. Lifshitz, Zh. Eksp. Teor. Fiz. 11 255 (1949).\\
$[31]$ R. G. Cai et al, Phys. Rev. D 80 041501 (2009).\\
$[32]$ G. Calcagni, JHEP 0909 112 (2009).\\
$[33]$ H. Lu et al, Phys. Rev. Lett. 103 091301 (2009).\\
$[34]$ V. Sahni, T. D. Saini, A. A. Starobinsky, and U. Alam, JETP
Lett. 77, 201 (2003).\\
$[35]$ V. Sahni, A. Shafieloo and A. A. Starobinsky, Phys. Rev. D
78, 103502 (2008). \\
$[36]$ J. Lu, L. Xu, Y. Gui and B. Chang, arXiv:0812.2074v2
[astro-ph];\\
$[37]$ R. L. Arnowitt, S. Deser and C. W. Misner, {\it The
Dynamics of General Relativity} appeared as Chapter 7, pp.
227-264, in {\it gravitation: an introduction to current
research}, L. Witten, ed. (Wiley, New York, 1962), arXiv:
gr-qc/0405109.\\
$[38]$ C. Charmousis, G. Niz, A. Padilla and P. M. Saffin, {\it
JHEP} {\bf 0908} 070 (2009); T. P. Sotiriou, M. Visser and S.
Weinfurtner, JHEP 0910 (2009) 033; C. Bogdanos and E. N.
Saridakis, {\it Class. Quant. Grav.} {\bf 27} 075005 (2010).\\

\end{document}